\documentclass[preprint,showpacs,amsmath,amssymb,aps]{revtex4}
\usepackage{graphicx}
\begin{document}

\title{Remarks on  monopole charge properties within the Generalized Coherent State Model}

\author{A. A. Raduta$^{a),b),c)}$,  C. M. Raduta $^{b)}$, E. Moya de Guerra $^{c),d)}$ 
and P. Sarriguren $^{d)}$  }

\address{$^{a)}$ Department of Theoretical Physics and Mathematics, Bucharest
  University, POBox MG11, Romania}

\address{$^{b)}$ Department of Theoretical Physics, Institute of Physics and
  Nuclear Engineering, Bucharest, POBox MG6, Romania}

\address{$^{c)}$ Departamento de Fisica Atomica, Molecular y Nuclear, Universidad Complutense de Madrid, E-28040 Madrid, Spain}

\address{$^{d)}$ Instituto de Estructura de la Materia, CSIC, Serrano 123, E-28006 Madrid, Spain}

\begin{abstract}
The Generalized Coherent State Model, proposed previously for a unified description of magnetic and electric collective properties of nuclear systems, is used to study the ground state band charge density as well as the E0 transitions from $0^+_{\beta}$ to $0^+_g$.
The influence of the nuclear deformation and of angular momentum projection  on the  charge density is investigated. The monopole transition amplitude has been calculated  for ten nuclei. The results are compared with some previous theoretical studies and with the available experimental data. Our results concerning angular momentum projection are consistent with those of 
previous microscopic calculations for the ground state density. The calculations for the E0 transitions agree quite well with the experimental data. Issues like how the shape transitions or shape coexistence are reflected in the $\rho(E0)$ behavior 
are also addressed.
\end{abstract}
\maketitle

\renewcommand{\theequation}{1.\arabic{equation}}
\setcounter{equation}{0}

\section{Introduction}
\label{sec:level 1}
One body transition density operators play an important role in the microscopic description of various properties showing up in nuclear systems. For example, the charge density operator matrix elements corresponding to the ground state of a spherical system can be determined, with high precision,  in  elastic electron scattering, which results in having precious information about the spatial charge distribution. Similarly, the matrix elements between the ground state and excited states within the ground band might provide information about the nuclear shape \cite{Coop}. Indeed, in a electron scattering experiment at large momentum transfer the radial dependence of the charge distribution can be directly measured. Combining this result with other information on electromagnetic interaction in the considered nuclei, in the reference quoted above more refined statements on the deformed shapes could be made. 

The structure of the spectra in deformed nuclei requires the use of a deformed mean field. The final state describing an interacting many body system is a deformed state and, therefore, its use for the description of the transition probabilities requires the projection of the components with good angular momentum.  In particular, in order to account for the contribution of the tensorial components of the charge density, the many body ground state built up with deformed single particle states must be projected over the angular momentum \cite{Zari}.  There is no doubt that projection is of paramount importance for transitional and deformed nuclei. However, the correction brought by projection to the results obtained in the intrinsic frame  depends on the observable under consideration \cite{Vill,Elvi}. For example,  averaging a model Hamiltonian on an intrinsic ground state yields the system energy, while averaging it on  angular momentum projected states, a whole band of energy levels is obtained. 
For instance, for the collective magnetic dipole state with a deformed single particle basis one obtains $K=1$ states which are abusively called $1^+$ states. To our knowledge there is no rigorous proof that the admixture with the components of angular momentum  2 is negligible. Another example we want to comment upon, is that of the rotational bands which are considered to be a set of states characterized by the same quantum number $K$. However, in the laboratory frame of reference, $K$ is not a good quantum number and the meaning of a $K$-state is actually a state with a dominant $K$-component. The effect of projection is
felt by the operator matrix elements. There are cases of operators whose matrix elements are affected very little by the the angular momentum projection of the intrinsic states. The simplest case is the one when the operator is just a C-number 
constant.
Its matrix elements in the unprojected and projected states are equal to each other. At first glance, that would suggest that other operators insensitive to projecting the angular momentum from the intrinsic wave function, would be scalar operators. Of course, that is not true and an example is the boson number operator in a phenomenological picture. 

One issue of the present paper is to study the scalar part of the charge density operator within the generalized coherent state model (GCSM). Thus, we shall consider the matrix element of the charge density operator, truncated at its scalar term, on the unprojected ground state and, alternatively, on the projected $J^{\pi}=0^+$ state. We address also the question of how different are these matrix elements from those corresponding to a high angular momentum projected state. All matrix elements quoted above are studied as function of nuclear deformation.  

Another scalar operator which is studied here is the E0 transition operator. The monopole transition is often used to characterize various states of angular momentum equal to zero. Thus, in Ref.\cite{Sarr} two of us (P.S. and E.M.) expressed the monopole for the transition $0^+_{\beta}\to 0^+_g$ strength in terms of the mixing coefficient of the two states characterized by different deformations but lying close to each other in energy. In this way the transition strength may provide the mixing coefficient for the two states. In $^{158}$Gd several $0^+$ states have been seen in a (p,t) experiment \cite{Les}. These states  have been microscopically interpreted within a projected shell model and, alternatively, within the quasiparticle-phonon model \cite{Nic}. The authors of Ref.\cite{Nic} calculated the E2 strength for the transition from the ground state to the first $2^+$ state, the E0 strength for the transition from the excited $0^+$ states to the ground state, and the two-nucleon spectroscopic amplitudes. The experimental strengths for E0 and E2 transitions are concentrated in one and two states respectively, while the theoretical results \cite{Nic} show a large fragmentation of the two transition strengths. The experimental spectroscopic amplitudes indicate that two states are mainly populated, which contrasts the theoretical result where the  amplitude is fragmented among several states. From the analysis of Ref.\cite{Nic} it seems that the E0 strength is a signature only for one excited $0^+$ state, as in fact was considered in Ref.\cite{Sarr}.

The project mentioned above will be described according to the following plan. In Section II a brief description of the Generalized Coherent State Model will be given. That will help us to introduce the useful notations and to approach a self-contained presentation. The charge density expansion in terms of the quadrupole collective coordinates is given in Section III.
The monopole transition amplitude is treated in Section IV. The numerical applications are given in Section V, while
the final conclusions are summarized in Section VI.

\renewcommand{\theequation}{2.\arabic{equation}}
\setcounter{equation}{0}

\section{The Generalized Coherent State Model}
\label{sec:level 2}
The description of magnetic properties in nuclei has always been a central
issue. The reason is that the two systems of protons and neutrons respond
differently when they interact with an external electromagnetic field.
Differences are due to the fact that by contrast to neutrons,  protons are
charged particles, the proton and neutron magnetic moments are different from
each other and, finally, the proton and neutron numbers in a given nucleus are, in general,  different.

Many papers have been devoted to explaining  various features of the collective
dipole mode called, conventionally, scissors mode. The name of the mode  was suggested by Lo Iudice and Palumbo
who interpreted the dipole mode, within the Two Rotor Model \cite{Iudi}, as a scissors like oscillation of proton and
neutron systems described by two axially symmetric ellipsoids, respectively.

The Coherent State Model (CSM), proposed by Raduta {\it et al.} to describe the lowest three collective interacting bands \cite{Rad1},  was extended  by including the isospin degrees of freedom in order to
 account for the collective  properties of the scissors mode \cite{Rad2}.
This extension is conventionally called ``The Generalized Coherent State Model''(GCSM).

CSM starts with the construction of a restricted collective space, by projecting out the components of good angular momentum from three orthogonal quadrupole boson states. These states are chosen such that they are orthogonal before and after 
projection.
One of the three deformed states, the intrinsic ground state, is a coherent state of Glauber type with respect to the zero component of the quadrupole boson, $b^{\dagger}_{20}$, while the other two are obtained by acting with elementary boson polynomials on the ground state. In choosing the intrinsic excited states we take care that the projected states considered 
 in the vibrational limit have to provide the multi-phonon vibrational spectrum,  while for the large deformation regime their behavior coincides with that predicted by the liquid drop model.   
 
In contrast to the CSM, which uses only one boson for the composite system
of protons and neutrons, within the GCSM the protons are described by quadrupole
proton-like bosons, $b^{\dagger}_{p\mu}$, while the neutrons by quadrupole neutron-like bosons, $b^{\dagger}_{n\mu}$ .
Since one deals with two quadrupole bosons instead of one, one expects 
to have a more flexible model and to find a simpler solution satisfying the restrictions
required by CSM.  The restricted  collective space is defined  by the states describing the three
major bands, ground, beta and gamma, as well as the band  based on
the  isovector state $1^+$. Orthogonality conditions are satisfied by the
following 6 functions which generate by angular momentum projection,
6 rotational bands:
\begin{eqnarray}
\phi^{(g)}_{JM}&=&N^{(g)}_JP^J_{M0}\psi_g,~~
\psi_g=exp[(d_pb^{\dag}_{p0}+d_nb^{\dag}_{n0})-(d_pb_{p0}+d_nb_{n0})]
|0\rangle,
\nonumber\\
\phi^{(\beta)}_{JM}&=&N^{(\beta)}_JP^J_{M0}\Omega_{\beta}\psi_g,
\nonumber\\
\phi^{(\gamma)}_{JM}&=&N^{(\gamma)}_JP^J_{M2}(b^{\dag}_{n2}-b^{\dag}_{p2})\psi_g,
\nonumber\\
\tilde{\phi}^{(\gamma)}_{JM}&=&\tilde{N}^{(\gamma)}_JP^J_{M2}(\Omega^{\dag}_{\gamma,p,2}+\Omega^{\dag}_{\gamma,n,2})\psi_g,
\nonumber\\
\phi^{(1)}_{JM}&=&N^{(1)}_JP^J_{M1}(b^{\dag}_nb^{\dag}_p)_{11}\psi_g,
\nonumber\\
\tilde{\phi}^{(1)}_{JM}&=&\tilde{N}^{(1)}_JP^J_{M1}(b^{\dag}_{n1}-b^{\dag}_{p1})\Omega^{\dag}_{\beta}\psi_g.
\label{figcsm}
\end{eqnarray}
Here, the following notations have been used:
\begin{eqnarray}
\Omega^{\dag}_{\gamma,k,2}&=&(b^{\dag}_kb^{\dag}_k)_{22}+d_k\sqrt{\frac{2}{7}}
b^{\dag}_{k2},~~k=p,n,
\nonumber\\
\Omega^{\dag}_{\beta}&=&\Omega^{\dag}_p+\Omega^{\dag}_n-2\Omega^{\dag}_{pn},
\nonumber\\
\Omega^{\dag}_k&=&(b^{\dag}_kb^{\dag}_k)_0-\sqrt{\frac{1}{5}}d^2_k,~~k=p,n,
\nonumber\\
\Omega^{\dag}_{pn}&=&(b^{\dag}_pb^{\dag}_n)_0-\sqrt{\frac{1}{5}}d^2_p,
\nonumber\\
\hat{N}_{pn}&=&\sum_{m}b^{\dag}_{pm}b_{nm},~\hat{N}_{np}=(\hat{N}_{pn})^{\dag},~~
\hat{N}_k=\sum_{m}b^{\dag}_{km}b_{km},~k=p,n.
\label{omegagen}
\end{eqnarray}
Note that apriory we cannot select one of the two sets of states
$\phi^{(\gamma)}_{JM}$ and $\tilde{\phi}^{(\gamma)}_{JM}$ for gamma band, although  one is symmetric and the other asymmetric against proton neutron permutation.
The same is true for the two isovector candidates for the dipole states.
In Ref.\cite{Rad3}, results obtained by using alternatively a symmetric and an asymmetric structure
for the gamma band states were presented. Therein it was shown that the asymmetric structure
for the gamma band does not conflict any of the available data. By contrary,
considering for the gamma states an asymmetric structure and fitting the model
Hamiltonian coefficients in the manner described  in Ref.\cite{Rad3}, a better
description for the beta band energies is obtained. Moreover, in that situation
the description of the E2 transition becomes technically very simple.
For these reasons, here we make the option for a proton neutron asymmetric
gamma band.

All calculations performed so far considered equal deformations for protons and neutrons. The deformation parameter for the composite system is:
\begin{equation}
d=\sqrt{2}d_p=\sqrt{2}d_n.
\end{equation}
The factors $N$ involved in the expressions of wave functions are normalization constants calculated in terms of some overlap integrals.

We seek now an effective Hamiltonian for which the projected states (\ref{figcsm}) are, at least in a good approximation, eigenstates in the restricted collective space.
The simplest Hamiltonian fulfilling this condition is:
\begin{eqnarray}
H&=&A_1(\hat{N}_p+\hat{N}_n)+A_2(\hat{N}_{pn}+\hat{N}_{np})+
\frac{\sqrt{5}}{2}(A_1+A_2)(\Omega^{\dag}_{pn}+\Omega_{np})
\nonumber\\
&&+A_3(\Omega^{\dag}_p\Omega_n+\Omega^{\dag}_n\Omega_p-2\Omega^{\dag}_{pn}
\Omega_{np})+A_4\hat{J}^2.
\label{HGCSM}
\end{eqnarray}
The Hamiltonian given by Eq.(\ref{HGCSM}) has  only one off-diagonal matrix element in the basis (\ref{figcsm}). That is $\langle\phi^{\beta}_{JM}|H|\tilde{\phi}^{(\gamma)}_{JM}\rangle$.
However, our calculations show that this affects the energies of $\beta$ and $\tilde{\gamma}$ bands by an amount of a few keV. Therefore, the excitation energies of the six bands are in a very good approximation given by the diagonal element:
\begin{equation}
E^{(k)}_J=\langle\phi^{(k)}_{JM}|H|\phi^{(k)}_{JM}\rangle-
\langle\phi^{(g)}_{00}|H|\phi^{(g)}_{00}\rangle,\;\;k=g,\beta,\gamma,1,\tilde{\gamma},\tilde{1}.
\label{EkJ}
\end{equation}
It can be easily checked that the model Hamiltonian does not commute with the components of the $\hat{F}$ spin operator:
\begin{equation}
\hat{F}_0=\frac{1}{2}(\hat{N}_p-\hat{N}_n),\;\hat{F}_+=\hat{N}_{pn},\;\hat{F}_-=\hat{N}_{np}.
\label{Fspin}
\end{equation}

Hence, the eigenstates of $H$ are $F_0$ mixed states. However, the expectation
values of the $F_0$ operator on the projected model states are equal to zero.
This is caused by the fact that the proton and neutron deformations are considered
to be equal. In this case the states are of definite parity, with respect to
the proton-neutron permutation, which is consistent with the structure of the
model Hamiltonian which is invariant with respect to such a symmetry
transformation. To conclude, by contrast to the IBA2 Hamiltonian, the GCSM
Hamiltonian is not $\hat{F}$ spin invariant. Another difference to the IBA2, the most
essential one, is that the GCSM Hamiltonian does not commute with the boson number
operators. Due to this feature the coherent state approach proves to be the
most adequate one to treat the Hamiltonian in Eq.(2.4).
The asymptotic behavior of the magnetic state $1^+$, derived in Ref.\cite{Rad2},
shows clearly that
the phenomenological description of two liquid drops and two rigid rotors are
just particular cases  of GCSM, defined by specific restrictions.

The GCSM seems to be the only phenomenological model which treats simultaneously the
M1 and E2 properties. Indeed, in Refs.\cite{Rad3,Rad4} the ground, beta and gamma bands
are considered
together with a $K^{\pi}=1^+$ band built on the top of the scissor mode $1^+$.
By contrast to the other phenomenological and microscopic models, which treat
the scissors mode in the intrinsic reference frame, here one deals with states of good
angular momentum and, therefore,
there is no need to restore the rotational symmetry.
As shown in Ref.\cite{Iud}, the GCSM provides
for the total M1 strength an expression which is proportional to the
nuclear deformation
squared. Consequently, the M1 strength of $1^+$ and the B(E2) value
for $2^+$ are
proportional to each other, although the first quantity is determined by the
convection current
while the second one by the static charge distribution.

One weak point of most phenomenological models is that they use expressions
for transition operators not consistent with the structure of the
model Hamiltonian. Thus, the transition probabilities are influenced by the
chosen Hamiltonian only through the wave functions.
By contradistinction, in Refs. \cite{Rad3,Rad4} the E2 transition operator, as well as the M1 form-factor are derived analytically by
using the equation of motion of the collective coordinates determined by the model Hamiltonian. In this way a consistent description of electric and magnetic properties of many nuclei was attained.

Here we complete the GCSM achievements by considering scalar operators such as those involved in the ground state charge density and the monopole 
transitions, $\rho(E0)$. As we have already mentioned we address issues like, how sensitive are these quantities to the projection operation and also what is the influence of  the nuclear deformation $^{[1]}$. \setcounter{footnote}{1}\footnotetext{ Note that in our previous publications the total deformation was denoted by $\rho$. Here $\rho$ stands for the charge density while the nuclear deformation for the composite system is denoted by $d$.}

\renewcommand{\theequation}{3.\arabic{equation}}
\setcounter{equation}{0}

\section{The charge density in the liquid drop model}
\label{sec:level 3}
Suppose that the nuclear charge is distributed uniformly inside the nuclear surface described by: 
$\alpha_{2\mu}$:
\begin{equation}
R(\theta,\varphi)=R_0\left(1+\sum_{\lambda=0,2;\mu}\alpha_{\lambda\mu}^*Y_{\lambda\mu}\right)\equiv R_0+\Delta R,
\end{equation}
with $\alpha_{\lambda\mu}$ collective coordinates to be quantized later on.
The charge density has the expression:
\begin{equation}
\rho(r,\theta,\varphi)=\rho_0H\left[R(\theta,\varphi)-r\right],
\end{equation}
where $H$ denotes the Heaviside function while $\rho_0$ is the constant density corresponding to a sphere of radius 
$R_0=1.2A^{1/3} fm$. Expanding the charge density around the surface corresponding to vanishing quadrupole coordinates one obtains:
\begin{equation}
\rho(r,\theta,\varphi)=\rho_0\left[H(R_0-r)+\Delta R\delta(R_0-r)-\frac{1}{2}\left(\Delta R\right)^2\delta^{\prime}(R_0-r)+...\right].
\end{equation}
In  momentum space the charge density can be written  as a sum of tensor operators of various ranks. For example the term of rank $\lambda$ and projection $\mu$ reads:
\begin{equation}
\rho_{\lambda\mu}(q)={\cal C}\int r^2j_\lambda(qr)\left(\int \rho(r,\theta,\varphi)Y_{\lambda\mu}d\Omega\right)dr.
\end{equation}
Here $j_{\lambda}$ is the spherical Bessel function of first kind. The transfer momentum, during the scattering process with a charged particle, is denoted by $q$. ${\cal C}$ is a normalization factor which might be chosen such that for $q=0$ the density $\rho_0 $ is obtained. Here we choose ${\cal C}=1$, which means that in momentum space we deal with the total charge instead of charge density.

Let us consider first the scalar term involved in the expression of the charge density.
Taking into account the fact that the volume conservation restriction yields a relation between the monopole and quadrupole coordinates,

\begin{equation}
\alpha_{00}=-\frac{1}{\sqrt{4\pi}}\sum_{\mu}|\alpha_{2\mu}|^2,
\end{equation}
one obtains:
\begin{equation}
\rho_{00}(q)=\frac{3Ze}{qR_0}j_1(qR_0)-\frac{3}{8\pi}ZeqR_0j_1(qR_0)\sum_{\mu}|\alpha_{2\mu}|^2.
\label{ro0}
\end{equation}
Quantizing the quadrupole collective coordinate we can define the transition monopole operator, $\hat{\rho}_{00}$.
The elastic monopole form-factor is obtained as the expectation value of $\hat{\rho}_{00}$ on the ground state  wave 
function in the collective space. Here we consider alternatively the unprojected ground state and  the projected states describing the $J$-members of the ground band. In order to calculate the expectation values of the monopole charge density operator
in the states mentioned above, we have to express the coordinates in terms of boson operators through the canonical transformation:
\begin{eqnarray}
\hat{\alpha}_{2\mu}&=&\frac{1}{k_p\sqrt{2}}\left(b^{\dagger}_{p\mu}+(-)^{\mu}b_{p-\mu}\right),\nonumber\\
\pi_{2\mu}   &=&\frac{ik_p}{\sqrt{2}}\left((-)^{\mu}b^{\dagger}_{p-\mu}-b_{p\mu}\right).
\end{eqnarray}
The transformation relating the coordinates and conjugate momenta with the boson operators $b^{\dagger}_{p\mu}, b_{p\mu}$, is determined up to a multiplicative constant, $k_p$. This is at our disposal and will be fixed in several alternative ways described along this chapter.
 The results for the average values are:
\begin{eqnarray}
\langle \psi_g|\sum_{\mu}|\hat{\alpha}_{2\mu}|^2|\psi_g\rangle& =&\frac{1}{k_p^2}\left(d^2+\frac{5}{2}\right),\nonumber\\
\langle \phi^g_{JM}|\sum_{\mu}|\hat{\alpha}_{2\mu}|^2|\phi^g_{JM}\rangle& =&\frac{1}{k_p^2}\left(\frac{d^2}{2}+\frac{5}{2}\right)+
\frac{d^2}{2k_p^2}\frac{I^{(1)}_J(d^2)}{I^{(0)}_J(d^2)},
\end{eqnarray}
with

\begin{eqnarray}
I^{(0)}_J(y)&=&\int_{0}^{1}P_J(x)e^{yP_2(x)}dx,\nonumber\\
I^{(1)}_J(y)&=&\frac{\partial I^{(0)}_J(y)}{\partial y},\nonumber\\
        y&=&d^2,\;\;d=\sqrt{2}d_p.
\end{eqnarray}
In the above expressions $P_J(x)$ denotes the Legendre polynomial of rank $J$. Denoting by:
\begin{eqnarray}
A(q)&=&\frac{3Ze}{qR_0}j_1(qR_0),\nonumber\\
C(q)&=&-\frac{3}{8\pi}ZeqR_0j_1(qR_0),
\label{AC}
\end{eqnarray}
the matrix elements of the charge operator read:
\begin{eqnarray}
\langle \psi_g|\hat{\rho}_{00}(q)|\psi_g\rangle& =&A(q)+\frac{1}{k_p^2}\left(d^2+\frac{5}{2}\right)C(q),\\
\langle \phi^{(g)}_{JM}|\hat{\rho}_{00}(q)|\phi^{(g)}_{JM}\rangle& =&A(q)+C(q)\left[\frac{1}{2k_p^2}\left(d^2+5\right)+
\frac{d^2}{2k_p^2}\frac{I^{(1)}_J(d^2)}{I^{(0)}_J(d^2)}\right].
\label{meofro}
\end{eqnarray}
These expressions correspond to $Ze$ times the elastic form factor in the intrinsic and laboratory frame, respectively. 
 we refer to them as total charge $Q$.
The above defined integrals (3.9) have been studied analytically in Refs.\cite{Rad5,Rad6,Rad7,Iud}. From the results obtained in the quoted papers one easily obtains simple expressions for the extreme regimes of near spherical and rotational behaviors.
The results for the case of $J=0$ state are:
\begin{eqnarray}
\langle \phi^{(g)}_{00}|\sum_{\mu}|\hat{\alpha}_{2\mu}|^2|\phi^{(g)}_{00}\rangle& =&\frac{1}{2k_p^2}\left[\frac{d^4}{5}+d^2+5\right],\;\;
 d=\rm{small}\; (d\le 1).\nonumber\\
\langle \phi^{(g)}_{00}|\sum_{\mu}|\hat{\alpha}_{2\mu}|^2|\phi^{(g)}_{00}\rangle& =&\frac{1}{k_p^2}\left[d^2+2-\frac{2}{9}\frac{1}{d^2}\right],\;\;d=\rm{large} (d\ge 3).
\end{eqnarray}
We recall that for well deformed nuclei $d$ is typically greater than 3.
In the low momentum regime ($qR_0<<1$),  the expression (\ref{ro0}) is  much simplified:
\begin{equation}
\rho_{00}(q)=Ze\left[1-\frac{1}{10}\left(qR_0\right)^2-\frac{1}{8\pi}\left(qR_0\right)^2\sum_{\mu}|\alpha_{2\mu}|^2\right].
\end{equation}
Let us turn our attention to the quadrupole component of the charge density. Following the same procedure as in the case of the monopole component, we obtain:
\begin{equation}
\rho_{2\mu}=\int r^2j_2(qr)\left[\int \rho(r,\theta,\varphi)Y_{2\mu} d\Omega\right]dr=\rho_0R^3_0j_2(qR_0)\alpha_{2\mu}.
\end{equation}
Under the restriction $qR_0<<1$, the result is:
\begin{equation}
\rho_{2\mu}=\frac{3Ze}{40\pi}\left(qR_0\right)^2\alpha_{2\mu}.
\end{equation} 
Concluding, in the second order expansion in the surface coordinates, the charge density is:
 
\begin{eqnarray}
\rho_{\mu}(q) &=&\frac{3Ze}{qR_0}j_1(qR_0)-\frac{3}{8\pi}Ze(qR_0)j_1(qR_0)\sum_{\mu}|\alpha_{2\mu}|^2
\nonumber\\
&+& \frac{3Ze}{4\pi}j_2(qR_0)\alpha_{2\mu}.
\label{ro}
\end{eqnarray}
Thus, $\rho$ is expressed as a second order polynomial in $\alpha$: 
\begin{equation}
\rho_{\mu}(q)=A(q)+B(q)\alpha_{2\mu}+C(q)\sum_{\mu}|\alpha_{2\mu}|^2,
\end{equation} 
with the coefficients depending on the transferred momentum, defined by Eq. (\ref{AC}) and
\begin{equation}
B(q)= \frac{3Ze}{4\pi}j_2(qR_0) .
\label{C}
\end{equation}

In the intrinsic reference frame the expression becomes:
\begin{equation}
\rho_{\mu}(q)=A(q)+B(q)\left(\delta_{\mu , 0}\beta\cos\gamma +(\delta_{\mu,2}+\delta_{\mu,-2})\frac{\beta\sin\gamma}{\sqrt{2}}
\right)+C(q)\beta^2.
\end{equation}
We notice that the surface of constant charge density is of an ellipsoidal form which is consistent with the liquid drop shape. Coupling a particle to such a core system, the single particle motion would be determined by a quadrupole deformed mean field. 
In the boson representation, defined above,  one obtains:
\begin{equation}
\hat{\rho}_{\mu}(q)=A(q)+\frac{5C(q)}{2k_p^2}+\frac{B(q)}{k_p\sqrt{2}}\left(b^{\dagger}_{p\mu}+(-)^{\mu}b_{p,-\mu}\right)
+\frac{C(q)}{k_p^2}\hat{N}_p+\frac{C(q)}{2k_p^2}\left(b^{\dagger}_{p\mu}b^{\dagger}_{p,-\mu}+b_{p,-\mu}b_{p\mu}\right)(-)^{\mu},
\end{equation}
where $\hat{N}_p$ denotes the proton boson number operator.
The boson term $\left(b^{\dagger}_{p\mu}b^{\dagger}_{p-\mu}+b_{p-\mu}b_{p\mu}\right)(-)^{\mu}$ has diagonal matrix elements in ground and beta bands much larger than the off-diagonal one. Moreover, the matrix elements do not depend on the angular 
momenta of the states involved. For this reason we shall replace it by its average value, which is equal to $2d_p^2$.
Under these circumstances the zero component of the charge density operator becomes:
\begin{equation}
\hat{\rho}_0(q)={\cal T}+\frac{B}{k_p\sqrt{2}}\left(b^{\dagger}_{p0}+b_{p0}\right)+\frac{C}{k_p^2}\hat{N}_p,
\end{equation}
where
\begin{equation}
{\cal T}=A+\frac{C}{2k_p^2}(d^2+5).
\end{equation}
Acting with this boson operator on the unprojected ground state, one obtains:
\begin{equation}
\hat{\rho}_0(q)\psi_g=\left[\left({\cal T}+\frac{B}{\sqrt{2}k_p}d_p\right)+\left(\frac{B}{\sqrt{2}k_p}+\frac{Cd_p}{k_p^2}\right)b^{\dagger}_{p0}\right]\psi_g.
\end{equation}
We recall the fact that the canonical transformation relating the quadrupole coordinate and conjugate momenta with the boson operators is determined up to a multiplicative constant which was denoted by $k_p$. Taking for this constant the value
\begin{equation}
k_p=-\frac{C}{B}d,
\label{kpeq}
\end{equation}
the unprojected ground state becomes eigenfunction for the boson operator $\hat{\rho}$:
\begin{equation}
\hat{\rho}_{0}(q)\psi_g=\left(A+\frac{B^2}{2C}\frac{5}{d^2}\right)\psi_g.
\end{equation}
Considering the low momentum expansion for the coefficients $A, B$ and $C$, this equation becomes:
\begin{equation}
\hat{\rho}_{0}(q)\psi_g=Ze\left[1-\frac{\left(qR_0\right)^2}{10}\left(1+\frac{1}{2\pi d^2}\right)\right]\psi_g.
\end{equation}
Under these circumstances the parameter $k_p$ has a very simple expression:
\begin{equation}
k_p=\frac{5}{2}d.
\end{equation}
Alternatively, the canonicity parameter could be determined in the following way.
The stability condition for the average value of $\hat{\rho}$ on the unprojected ground state against the variation of $d$ provides the following equation for the deformation parameter $d$:
\begin{equation}
2Cd+k_pB=0.
\label{eqfork}
\end{equation}
However, in our previous investigations $d$ has been fixed by fitting some energies in the ground band. We could keep those values for $d$ and use Eq. (\ref{eqfork}) to determine $k_p$. We remark that the value of $k_p$ obtained in this way is twice as much as the one given by Eq.(\ref{kpeq}). In this case the low momentum regime provides $k_p=5d.$

At this stage it is worth recalling the way the canonicity parameter $k_p$ was fixed within the GCSM model when the M1 and E2 properties were investigated. 
In the asymptotic regime, i.e. $d$ large, the ground band energies can be expressed as \cite{Rad3}:
\begin{equation}
E^g_J=\left[\frac{A_1+A_2}{6d^2}+A_4\right]J(J+1).
\end{equation}
Equating this expression with that given by the liquid drop model, one finds an
equation
relating the nuclear deformation with the parameter $\rho$:
\begin{equation}
\beta_0^2=\frac{\pi}{3.24}\frac{\hbar^2}{M_N}A^{-5/3}\left[\frac{A_1+A_2}{6d^2}+A_4\right]^{-1}.
\label{beta02}
\end{equation}
Identifying this deformation with the average value of the second order
invariant in
$\alpha$'s coordinates and subtracting  the zero point motion contribution,
one finds:
\begin{equation}
k_p=\frac{d}{\beta_0}.
\end{equation}
In Table I, the values of $\beta_0^{-1}$ are compared with those of $k_p/d$ given by Eq. (\ref{kpeq}). We notice that the two sets of data are quite close to each other. 

\begin{table}
\begin{tabular}{|c|ccccccccccc|}
\hline
   & $^{152}$Gd& $^{154}$Gd& $^{156}$Gd &$^{158}$Gd&  $^{160}$Gd & $^{154}$Sm &$^{164}$Dy
&$^{168}$Er&  $^{174}$Yb& $^{232}$Th & $^{238}$U\\
\hline
$\beta_0^{-1}$&1.471&1.320 &1.176& 1.158& 1.146& 1.129 & 1.140 & 1.242 & 1.261 & 1.247 & 1.142
\\
$k_p/d$&1.406&1.393 & 1.381 &1.368 & 1.355 &1.394 &1.329 &1.303 &1.264 &0.864 &0.820 \\
\hline
\end{tabular}
\caption{The values for the $k_p/d$ ratio calculated in two alternative ways: a) according to Eq. (3.30), as in Ref.
\cite{Rad3}. In this case the ratio is equal to   $\beta_0^{-1}$, given by Eq. (\ref{beta02}), and the resulting values are given in the first row; b) the ratio is given by Eq. (\ref{kpeq}) for a transfer momentum $q=0.54 fm^{-1}$. The corresponding values are listed in the second row. }
\end{table}

\renewcommand{\theequation}{4.\arabic{equation}}
\setcounter{equation}{0}

\section{Electric monopole transition}
\label{sec:level 4}
The scattering process where the colliding particle may be inside the target nucleus involves longitudinal momenta associated to the Coulomb field \cite{Bohr}:
\begin{equation}
{\cal M}(C\lambda,\mu)=\int\rho({\bf r})f_{\lambda}Y_{\lambda\mu}(\theta,\varphi)d\tau,
\end{equation} 
where $f_{\lambda}$ is  a function depending on the radial motion of the particle inside nucleus. If the monopole Coulomb momentum is expanded in powers of $r$, then the lowest order term giving rise to an intrinsic transition is proportional to $r^2$. Therefore, the monopole operator responsible for the transition with $\lambda\pi=0+$ is:
\begin{equation}
m(E0)=\int \rho({\bf r})r^2d\tau ,
\end{equation}
where $\rho$ denotes the electric charge density.
Expanding $\rho$ in terms of the liquid drop coordinates $\alpha_{2\mu}$, we obtain:
\begin{equation}
m(E0)=\rho_0R_0^5\left(\frac{4\pi}{5}+\sqrt{4\pi}\alpha_{00}+2\sum_{\mu}|\alpha_{2\mu}|^2\right).
\end{equation}
Using the volume conservation condition for the monopole coordinate $\alpha_{00}$, the final result for the monopole moment is:

\begin{equation}
m(E0)=\frac{3}{5}ZeR_0^2\left[1+\frac{5}{4\pi}\sum_{\mu}|\alpha_{2\mu}|^2\right].
\end{equation}
The matrix element of this operator gives the amplitude for the transition probability between the involved states. In particular, for the transition $J^+_{\beta}\to J^{+}_g$ we obtain:
\begin{equation}
\rho(E0)\equiv \langle \phi^{(\beta)}_{JM}|m(E0)|\phi^{(g)}_{JM}\rangle =\frac{3\sqrt{5}ZeR_0^2}{8\sqrt{2}\pi k_p^2}.
\label{rhoe0}
\end{equation}
Note that the amplitude for the monopole transition is not depending on the state angular momentum.
Moreover, the same expression is obtained if  the projected states are replaced by the unprojected ground and beta states,
 respectively.

In nuclei which exhibit shape coexistence, calculations of $E0$ transitions could provide a test for the mixing amplitudes of states with different deformations, defining the ground state \cite{Sarr}. For these cases,  $\rho(E0)$ can be expressed in terms of the mixing coefficient $\lambda$ and the difference between the r.m.s. associated to the states involved in the E0 transition, i.e. the ground state $0^+_{g}$ and the beta state $0^+_{\beta}$.

In what follows we shall show how the shape coexistence may be investigated within the GCSM approach.
First we show that the monopole transition can be expressed in terms of r.m.s. radii of beta and ground bands.
Indeed, using Eq.(4.4) for $m(E0)$ the r.m.s radii of the states from ground and beta bands are defined as:

\begin{eqnarray}
\langle r^2\rangle ^{g}_{J}&=&\frac{3}{5}ZeR_0^2\left[1+\frac{5}{4\pi}\langle \phi^{(g)}_{JM}| \sum_{\mu}|\alpha_{2\mu}|^2
|\phi^{(g)}_{JM}\rangle \right],\nonumber\\
\langle r^2\rangle ^{\beta}_{J}&=&\frac{3}{5}ZeR_0^2\left[1+\frac{5}{4\pi}\langle \phi^{(\beta)}_{JM}| \sum_{\mu}|\alpha_{2\mu}|^2|\phi^{(\beta)}_{JM}\rangle \right],
\label{r2gb}
\end{eqnarray}
Note that dividing the above expressions by Z, one obtains the charge radii in the states of angular momentum $J$.
Both matrix elements involved in Eq.(\ref{r2gb}) can be expressed by the expectation value of the boson number operator $\hat{N}$, in the state $J^+$ from the ground band:
\begin{eqnarray}
\langle r^2\rangle ^{g}_{J}&=&\frac{3}{5}ZeR_0^2\left[1+\frac{5}{8\pi k_p^2}\left(d^2+5+2\langle \phi^{(g)}_{JM}|\hat{N}
|\phi^{(g)}_{JM}\rangle \right)\right],\nonumber\\
\langle r^2\rangle ^{\beta}_{J}&=&\frac{3}{5}ZeR_0^2\left[1+\frac{5}{8\pi k_p^2}\left(d^2+7+2\langle \phi^{(g)}_{JM}
|\hat{N}|\phi^{(g)}_{JM}\rangle \right)\right].
\label{r2grb}
\end{eqnarray}
From these relations we obtain that the difference  of the beta and the ground band r.m.s. does not depend on the angular momentum $J$.  Moreover,  the mentioned difference is related to $\rho(E0)$ by a very simple equation:
\begin{equation}
 \rho(E0)=\sqrt{\frac{5}{8}}\left(\langle r^2\rangle ^{\beta}_0-\langle r^2\rangle ^{g}_0\right).
\label{roE0rbmg}
\end{equation} 

In Ref. \cite{Rad2} the projected states used by GCSM have been studied in the intrinsic reference frame and the result was that each state is a superposition of components with different quantum numbers $K$. However, the prevailing components have $K=0$ for ground and beta bands and $K=2$ for gamma band. Thus, the model is quite flexible for studying the band mixing. The question is whether the present formalism can be extended for studying the interaction between states of the same angular momenta and K. 
Indeed, GCSM can be used to  describe the collective properties of both gamma stable, where $E_{0^+_{\beta}}< E_{2^+_{\gamma}}$, and gamma unstable nuclei when the ordering of the had states of beta and gamma bands is opposite to the one mentioned above. For the gamma stable nuclei there are cases where the  state $0^+_{\beta}$ is low in energy. An attempting interpretation for such a situation assumes that this state belongs to the second well of the potential energy in the $\beta$ variable, while the ground state is located in the  well with a less deformed minimum. In what follows we shall show that our model is able to account for this kind of shape coexistence. Indeed, if the potential barrier is not high one can expect that the system is tunneling from one well to another and, therefore, is reasonable to assume that the real ground state is in fact a linear combination of the states 
$0^+_g$ and $0^+_{\beta}$. To simplify the notations hereafter the projected states with angular momentum zero from the ground and beta bands are denoted by $|0^+_g\rangle $ and $|0^+_{\beta}\rangle $, respectively.
Adding to the model Hamiltonian a term which couples the states from ground and beta bands, then the new Hamiltonian yields new eigenstates with angular momentum equal to zero:
\begin{eqnarray}
|0\rangle_I&=&\sqrt{\lambda}|0^+_g\rangle +\sqrt{1-\lambda}|0^+_{\beta}\rangle,\nonumber\\
|0\rangle_{II}&=&\sqrt{1-\lambda}|0^+_g\rangle -\sqrt{\lambda}|0^+_{\beta}\rangle.  
\end{eqnarray}
Using the above results one can calculate the amplitude of the $E0$ transition, relating the new states, i.e.
$0_{II}\to 0_{I}$. The final results is:
\begin{equation}
\rho _{I,II}(E0)=\left[-\sqrt{\lambda(1-\lambda)}+(1-2\lambda)\sqrt{\frac{5}{8}}\right]
\left(\langle r^2\rangle ^{\beta}_0-\langle r^2\rangle ^{g}_0\right).  
\label{rhoI,II}
\end{equation}
Replacing $\rho_{II,I}(E0)$  by the corresponding experimental value, the relation (\ref{rhoI,II}) becomes an equation for
the mixing coefficient $\lambda$ ,
\begin{equation}
-\sqrt{\frac{8\lambda(1-\lambda)}{5}}+(1-2\lambda)=F.
\label{eqforla}
\end{equation}
Here $F$ stands for the ratio between the experimental value for $\rho_{II,I}(E0)$ and the calculated value of $\rho(E0)$
given by Eq.(\ref{roE0rbmg}):
\begin{equation}
F=\frac{\rho^{exp}_{II,I}(E0)}{\rho(E0)}.
\end{equation}
Concluding, due to Eq.(\ref{rhoI,II}) the GCSM can provide information about the shape coexistence. 
On the other hand mixing states corresponding to different shapes may be used to improve the description of the E0 transitions.

\begin{figure}[ht!]
\begin{center}
\includegraphics[width=0.8\textwidth]{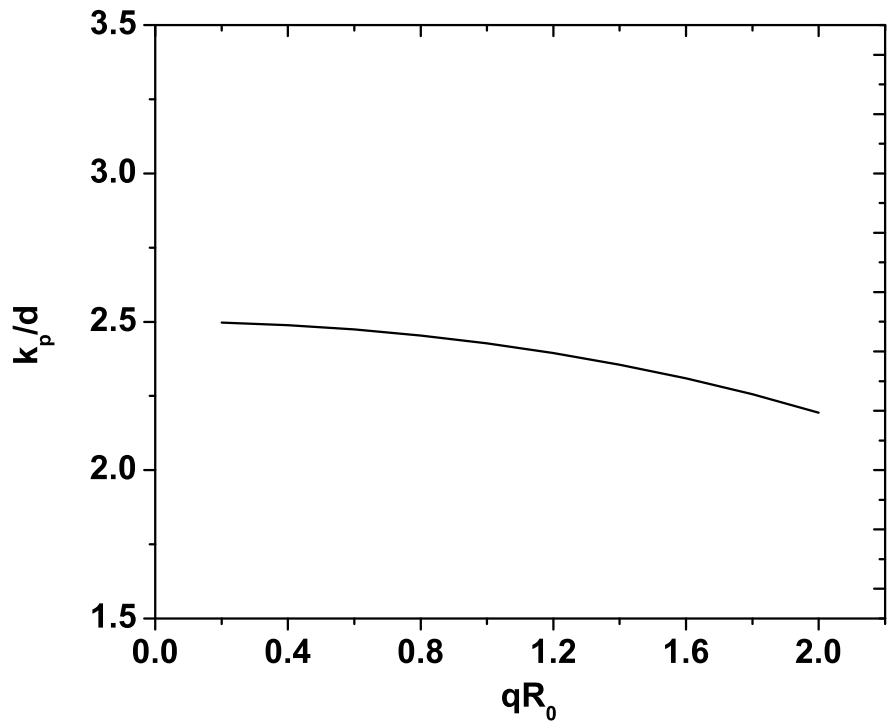}
\end{center}
\caption{The ratio $k_p/d$ with $k_p$, given by Eq.(\ref{kpeq}), is plotted as function of $qR_0.$ The curve corresponds to a deformation parameter $d=3.2195$ which, according to Ref.\cite{Rad3}, characterizes the isotope $^{156}$Gd.}
\label{Fig.1}
\end{figure}
\clearpage

\begin{figure}[ht!]
\begin{center}
\includegraphics[width=0.8\textwidth]{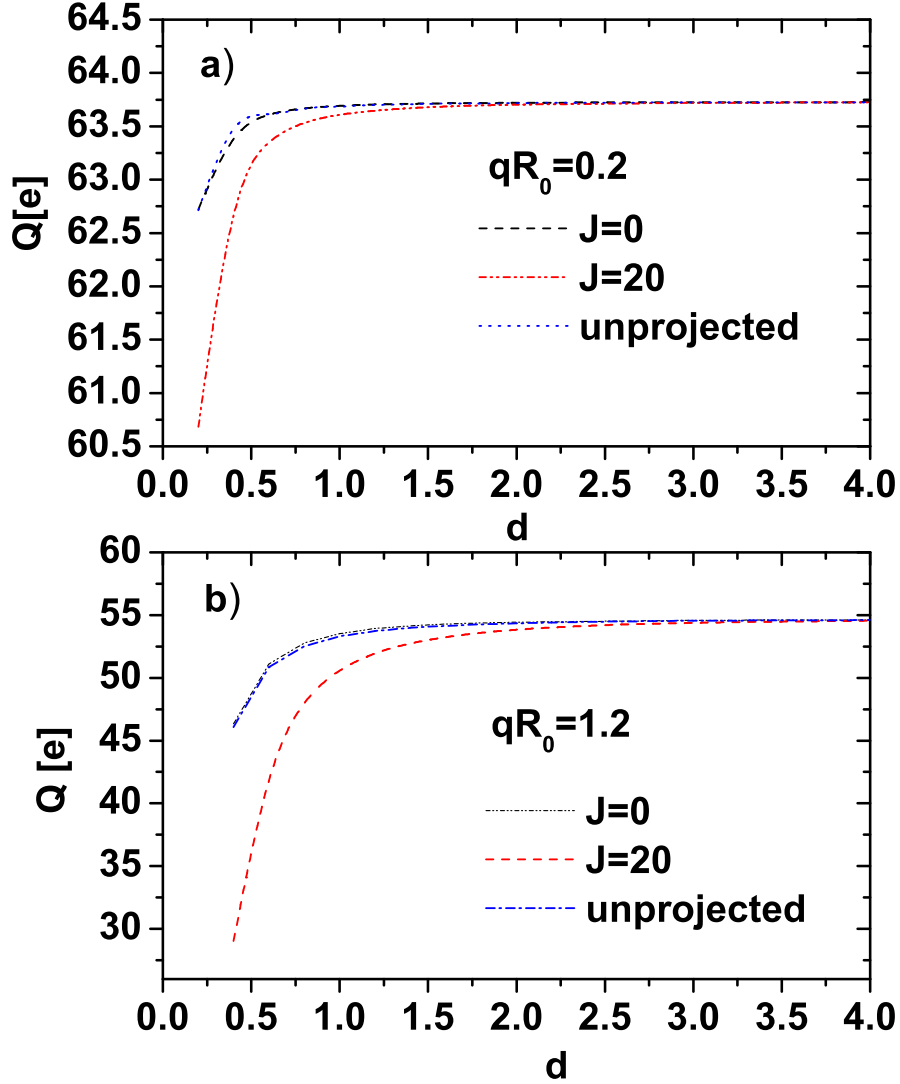}
\end{center}
\caption{Color online. The total charge of $^{156}$Gd is plotted as function of the nuclear deformation for  $qR_0=0.2$ (panel a)) and
$qR_0=1.2$ (panel b)). The results for unprojected ground state cannot be distinguished from those corresponding to the projected $J=0$ state. Moreover, for $d>1$ they are close to the nuclear charge Ze=64e if $qR_0=0.2$. For $qR_0=1.2$,  the charge for $J=20$ deviates from that corresponding to  $J=0$ and 
unprojected cases if $d<3$. The three curves converge to a common value which is close to  55e, at the end of interval. These calculations correspond to the value of $k_p$ given by Eq.(\ref{kpeq}). The matrix elements for unprojected and projected states have been calculated with  Eqs.(3.11), (\ref{meofro}), respectively. For the deformation parameter $d=3.2195$ predicted in Ref.\cite{Rad3} for $^{156}Gd$, the projection has no effect on the charge density.}
\label{Fig.2}
\end{figure}
\clearpage

\begin{figure}[ht!]
\begin{center}
\includegraphics[width=0.8\textwidth]{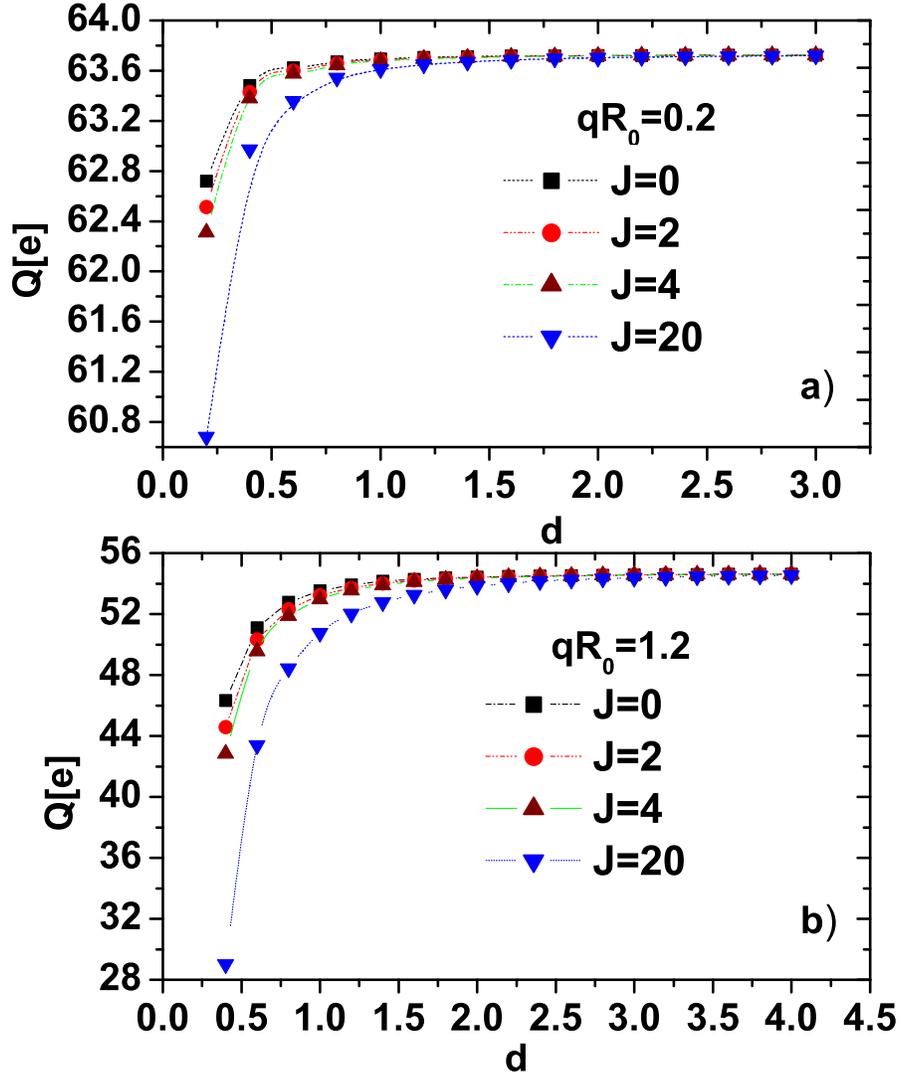}
\end{center}
\caption{Color online. The total charges of $^{154}$Gd in the ground band states with $J=0, 2, 4, 20$, respectively, are plotted as function of the nuclear deformation for 
$qR_0=0.2$ (panel a)) and $qR_0=1.2$ (panel b)). The results for unprojected ground state are almost the same as for the projected $J=0$ state and,
therefore, are not plotted here. These calculations correspond to $k_p$ given by Eq.(\ref{kpeq}). The charge for $J=20$ deviates from that corresponding to  $J=0,2,4$  
 if $d<3.$  For $d=3.0545$ corresponding to the chosen nucleus, the charges for the considered states are the same. However, a deviation of about ten units from the total charge is to be noticed for $qR_0=1.2$.}
\label{Fig.3}
\end{figure}
\begin{figure}[ht!]
\begin{center}
\includegraphics[width=0.8\textwidth]{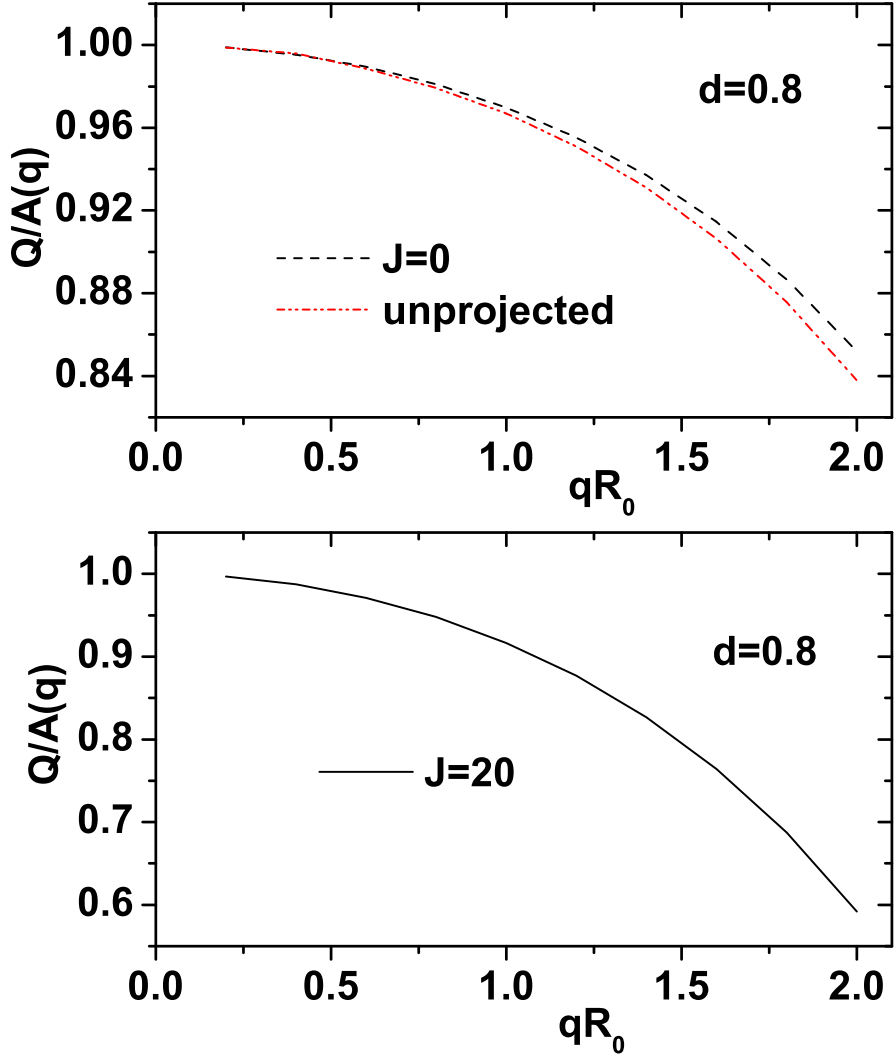}
\end{center}
\caption{Color online. The ratio between the total charge $Q$ and the term $A(q)$ in Eq.(3.11,12) is plotted as a function of $qR_0$ for $d=0.8$. In the upper panel the cases of unprojected ground state and of $J=0$ projected states are considered. In the bottom panel the case of projected $J=20$ state is presented. In both panels the parameter $k_p$ is calculated by means of Eq.(\ref{kpeq}). The nuclear radius $R_0$ corresponds to $^{156}$Gd.    }
\label{Fig.4}
\end{figure}
\begin{figure}[ht!]
\begin{center}
\includegraphics[width=0.8\textwidth]{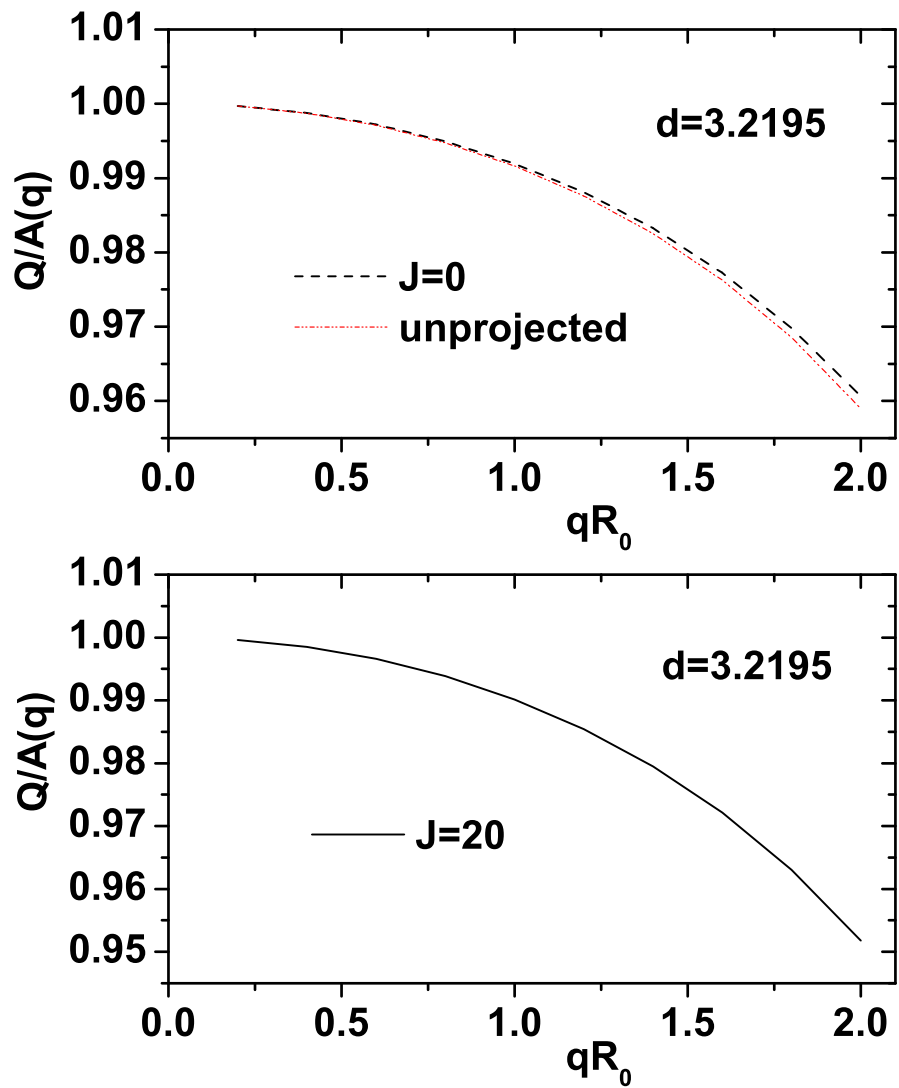}
\end{center}
\caption{Color online. The same as in Fig. 4 but for d=3.2195.   }
\label{Fig.5}
\end{figure}
\begin{figure}[ht!]
\begin{center}
\includegraphics[width=0.8\textwidth]{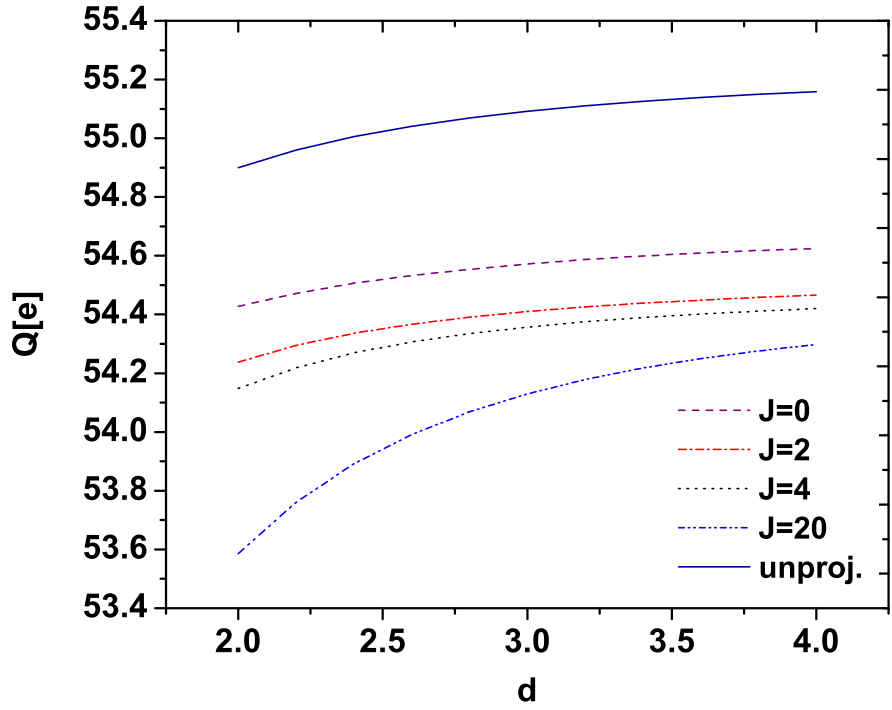}
\end{center}
\caption{Color online. The total charges of $^{156}$Gd in the unprojected ground state and the projected ground band states with J=0,2,4,20 respectively, are plotted as function of the nuclear deformation parameter d. Here the charge density expansion includes also the quadrupole term. The transfer momentum was taken as given by the equation: $qR_0=1.2.$}
\label{Fig.6}
\end{figure}

\clearpage
\renewcommand{\theequation}{5.\arabic{equation}}
\setcounter{equation}{0}
\section{Numerical results}
\label{sec:level 5}
Our numerical studies refer to the scalar term of the charge density as well as to the monopole transition from $0^{+}_{\beta}\to 0^{+}_g$. In both cases we intend to draw a definite conclusion about the effect of projection on these quantities. Also, we want to see how the projection effect depends on the nuclear deformation. 

As we have already seen in the previous Sections, the matrix elements of the charge density and monopole transition operator   depend on the parameter $k_p$. This parameter is proportional to the deformation parameter $d$. In Fig.1 we represent $k_p/d$ as a function of the product $qR_0$, where $R_0$ stands for the nuclear radius. In the interval $(0,2)$ for $qR_0$, the ratio is slowly decreasing from 2.5 to 2.2. Therefore, the value 2.5 obtained for $k_p/d$ in the low momentum regime could be considered as a reasonable approximation for the whole interval considered in Fig. 1.  

The matrix elements of the scalar part of the charge operator in the intrinsic and laboratory frame are given by Eqs. (3.11)
 and (\ref{meofro}), respectively. The latter gives the $q$ dependent charge of the system in the state $J^+$ of the ground band given by angular momentum projected state $\phi^{(g)}_{JM}$, while the former expresses the q dependent charge of the intrinsic ground state. In Fig. 2  the charge is represented as function of the deformation parameter $d$ for  $qR_0=0.2$ and $qR_0=1.2$ respectively. Calculations were made with $R_0$ corresponding to $^{156}$Gd. For both $qR_0$ values the charges of the system in the projected $J=0$ state and in the unprojected ground state are indistinguishable. As one increases the angular momentum, the effect of projection is larger, particularly at smaller $d$ values. The projection effect is vanishing for $d\geq 2.$ In the limit of $d\to 0$, the matrix elements (3.12) exhibit the behavior given by:
\begin{equation}
\langle \phi^{(g)}_{JM}|\hat{\rho}_{}|\phi^{(g)}_{JM}\rangle =A(q)+C(q)+\frac{J}{4d^2}\frac{B^2(q)}{C(q)}. 
\end{equation}
Due to this feature, for small deformations a large fall down of the curves corresponding to $J=20$ in Figs. 2,  is obtained.  
For $d>2$ the charges corresponding to the unprojected, the $J=0$ and the $J=20$ projected states, are about the same.
The common value of $Q$ is very close to the value 64, which is the nuclear charge of $^{156}$Gd. Also, the deformation parameter of $^{156}$Gd, determined in Ref.\cite{Rad3} to be 3.2195, lies on the saturation plateau. Actually this feature confirms that for deformed systems the strong coupling limit holds. 
The fact that for $qR_0=0.2$ the charge is close to the value 64, corresponds to the well known fact that the form factor is close to 1 when $q\ll 1/R_0$, and it is consistent with the assumption of fast convergence of the expansion of the charge density in terms of the quadrupole collective coordinates.
Similar features are seen in Fig.3 where the charge is calculated for $R_0$ corresponding to $^{154}$Gd. Here we added the results for $J=2$ and $J=4$ but we omitted those for the unprojected ground state since they are practically the same as for the projected ground state

Concluding, the projection operation does not affect the scalar $q$ dependent charge of deformed systems in the ground state.  A screening of charge for small deformation  and large angular momentum  is noticed. According to Fig. 2 b), for large $qR_0$ the screening shows up also for unprojected as well as for projected ground state. Moreover, for small deformation the deviation of the charge in the state with $J=20$  is substantially different from that corresponding to
the ground state. Similar features are seen in Fig. 3 b) for $^{154}$Gd.

In contrast to Ref.\cite{Zari}, here we deal with that part of the charge density which affects the elastic scattering
cross section. In the mentioned reference, the multipole  $\lambda $ terms of the charge density are essential in determining the ground to $\lambda^+_g$ excitation. However, these terms bring contributions also to the $J^+\to J^+$ matrix elements.
Indeed, considering the full expansion $\rho_0$ given by Eq.(3.18), the matrix elements for unprojected and projected states respectively, become:

\begin{eqnarray}
\langle \psi_g|\rho_{0}(q)|\psi_g\rangle& =&A(q)+\frac{d}{k_p}B(q)+\frac{1}{k_p^2}\left(d^2+\frac{5}{2}\right)C(q),\\
\langle \phi^{(g)}_{JJ}|\hat{\rho}_{00}(q)|\phi^{(g)}_{JJ}\rangle& =&A(q)+\frac{d}{k_p}B(q)C^{J~2~J}_{0~0~0}
C^{J~2~J}_{J~0~J}+C(q)\left[\frac{1}{2k_p^2}\left(d^2+5\right)+
\frac{d^2}{2k_p^2}\frac{I^{(1)}_J(d^2)}{I^{(0)}_J(d^2)}\right].\nonumber
\label{meofro2}
\end{eqnarray}
These matrix elements are plotted in Fig. 6 as functions of $d$. From this figure we see that the projection brings a correction of about 1\% to the total charge in the states $0^+, 2^+$ and  $4^+$, which is consistent with the microscopic calculations of Ref.\cite{Zari}. In the quoted reference a correction of about 15\% is noticed in the transition
$0^+\to 6^+$. Such a big correction is expected  to show up also in our formalism, if the multipole of rank 6 would be considered in the charge density expansion.

\begin{table}
\begin{tabular}{|c|cccc|}
\hline
 &\multicolumn{4}{c|}{$\rho(E0)[e.fm^2]$}\\ \cline{2-5}
 & Th.,$k_p=\frac{5}{2}d$ & Exp. &Th., $k_p=5d$ & Th., $k_p$ from ref. \cite{Rad3}\\
\hline
$^{152}$Gd&9.105   & 10.29$\pm$ 1.09 $^{d)}$& 2.276 & 26.284\\     
$^{154}$Gd& 8.574  & 11.749$\pm$ 0.101$^{a)}$& 2.143 & 30.765\\
          &        & 12.34 $\pm $1.13$^{d)}$ &       &       \\
$^{156}$Gd&7.784   & 7.469$\pm$ 0.071$^{a)}$& 1.946 & 35.170\\
          &        & 8.55 $\pm$ 0.96 $^{d)}$&       &       \\
$^{158}$Gd&6.286   & 5.487$\pm$ 0.465$^{b)}$& 1.571 & 29.282\\
          &        & 7.87 $\pm$ 1.25 $^{d)}$&       &        \\
$^{160}$Gd&5.606   &                               &1.401       & 26.688\\
$^{154}$Sm&7.116   &12.818$\pm$2.551$^{e)}$                               &1.779       & 34.912\\
$^{164}$Dy&5.804   &                               &1.451       & 27.341\\
$^{168}$Er&5.820   & 1.24$\pm$ 0.51$^{b,f)}$          &1.455                           & 23.567\\
$^{174}$Yb&6.083   & $< $1.85  $^{d)}$                    &1.521               & 23.906\\
$^{232}$Th&11.731  &13.646$\pm$ 4.88$^{b)}$        &2.933       & 47.142\\
$^{238}$U&11.228   & 5.502$\pm$ 0.05$^{c)}$        &2.807       & 53.847\\
         &         & 23.2$\pm$ 2.262$^{c)}$        &       &       \\
\hline
\end{tabular}
\caption{The monopole transition amplitudes predicted by Eq.(\ref{rhoe0}) are compared with the experimental data taken from Refs.\cite{Lang}$^{a)}$, \cite{Nic,Wood}
$^{b)}$, \cite{Gar}$^{f)}$, \cite{Cacsi}$^{c)}$, \cite{Kibe}$^{d)}$ and \cite{Wimm}$^{e)}$. Eq.(\ref{rhoe0}) has been used alternatively for 
$k_p$ for which the charge density operator admits the unprojected ground state as an eigenfunction (first column) and for $k_p$ which were used in Ref.\cite{Rad3} to describe the M1 and E2 properties of the nuclei listed in this Table}. 
\end{table}

\begin{figure}[ht!]
\begin{center}
\includegraphics[width=0.8\textwidth]{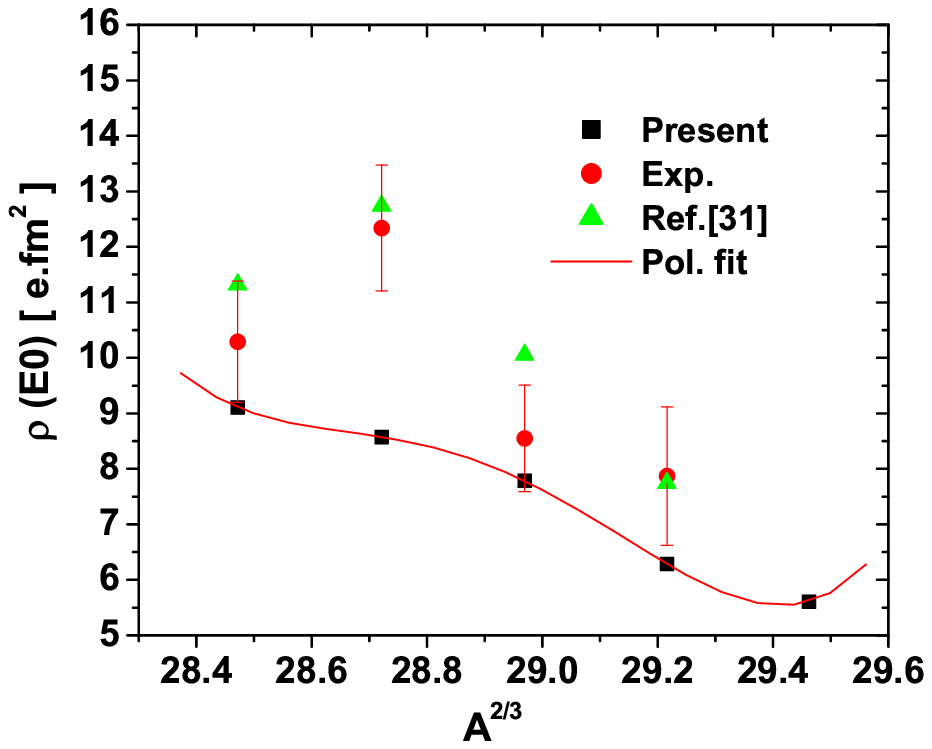}
\end{center}
\caption{Color online. The predicted values of $\rho(E0)$  for some even isotopes of $Gd$, are represented as function of $A^{2/3}$. For comparison, the experimental results are also given.}
\label{Fig.7}
\end{figure}

We note from Eqs.(3.11,12) that the  total charge is determined by summing two distinct terms, one depending exclusively on the transfer momentum, that is denoted by $A(q)$, and a  term which is a product of two factors depending on $q$ and $d$, respectively. One may ask oneself what is the relative contribution of these terms to the total charge for non-vanishing $q$ values. We addressed this question by studying the ratios $Q/A(q)$ as function of $qR_0$ for two values of the deformation parameter $d$. Thus, from Fig. 4 we see that the term depending on deformation may affect the charge of the ground state at most by 15\% for $qR_0=2$. In the state with J=20 and the quoted value of $qR_0$ the deformation relative contribution is of 40\%. For large nuclear deformation, d=3.2195, the relative contribution of the deformation is ranging from zero to 5\% when $qR_0$ is increased from 0 to 2. 

Now let us focus our attention on the monopole transition $0^+_{\beta}\to 0^+_g$. The transition amplitude was calculated with Eq.(\ref{rhoe0}). We notice that the monopole transition operator $m(E0)$ has an expression identical with that supplied by the liquid drop model. However, the wave functions are specific to GCSM and they may describe  the spherical and deformed nuclei in a unified way. Another feature which is specific to our description is the canonicity parameter $k_p$ defining the equations which relate the coordinates and conjugate momenta to the boson operators. Within the liquid drop model in its original form
the canonicity parameter is chosen such that the boson Hamiltonian does not contain a term like
$\sum_{\mu}(-)^{\mu}\left(b^{\dagger}_{2\mu}b^{\dagger}_{2-\mu}+b_{2-\mu}b_{2\mu}\right)$. This idea is not 
applicable to GCSM, since the starting Hamiltonian is anharmonic and, moreover, is considered in the boson picture. 
Here we present the results obtained by fixing $k_p$ in three different ways: a) From the minimum condition for the charge density and the low momentum transfer restriction. This provides a simple expression for  $k_p(=5d$); b) Requiring that the unprojected ground state is an eigenstate of the scalar part of the charge density operator. Note the fact that this condition is fulfilled automatically in microscopic models where a many body Slater determinant is eigenstate of the charge density operator. For a low momentum regime, the mentioned condition provides $k_p=\frac{5}{2}d$ ; c) As in Ref.\cite{Rad3} i.e., $k_p=\frac{d}{\beta_0},$ with $\beta_0$ fixed from the equation obtained by equating the expressions of the asymptotic energies in the ground band  and  that of the liquid drop model in the large deformation regime. The results of our calculations obtained with the three versions of fixing $k_p$ are given in Table II. The predictions are compared with the available experimental data for eleven nuclei. Notice that the data from the quoted references have been transformed by multiplying them with the factor $R_0^2$.

By inspection of Table II, we notice that except  for the cases of $^{168}$Er and $^{174}$Yb all the other data are reasonably well described by choosing $k_p=\frac{5}{2}d$. For $^{168}$Er and $^{174}$Yb, it seems that the version which provides $k_p=5d$ yields a good agreement with the corresponding experimental data. Using the parameter $k_p$ from Ref.\cite{Rad3}, which corresponds to a consistent description of the E2 and M1 properties, one obtains  $\rho(E0)$ values which exceed the experimental data by a factor ranging from 2.6 to 5. 

These discrepancies could be attributed to the fact that the collective coordinates respond differently to the interaction with longitudinal and transverse components of the electromagnetic external  field, respectively. The former components may determine a E0 excitation while the latter one can excite the nuclei
through, for example,  a $E2$ transition. 

The observation that different kappa's are needed to reproduce E2 and E0 properties reflects the fact that these are independent quantities. The E0 properties are not determined by the E2 ones and this perhaps suggests the necessity of introducing the monopole bosons. 

According to Table I the values of $k_p$ provided by the procedure labeled by c) are obtainable for a large transfer momentum, while those defined by a) and b) are obtained under the low momentum restriction.  The procedure c) might be suitable to fix the strength for  the quadrupole component of the charge density operator but not for the scalar component strength. An essential point in understanding these discrepancies is the fact that in Ref.\cite{Rad3} the quadrupole transition operator involves also the polarization effects of the neutron systems. Therefore the E2 transition is achieved by a combined contribution of proton and neutron systems. Here only the contribution of protons is considered and therefore the method c) of fixing $k_p$ is not adequate.

 We would like to mention that the model Hamiltonian used by GCSM (2.4) is a fourth order boson Hamiltonian, while the charge density expansion in collective quadrupole coordinates is truncated to second order. This lack of consistency  might be another source for the discrepancy between the values of $k_p$ obtained here and those given in Ref.\cite{Rad3}.

Note that the set of nuclei considered in the present paper involves a chain of even isotopes of Gd. Along this chain the shape undertakes a transition from  a spherical to a deformed one. The critical point of this transition is met in $^{154}$Gd
\cite{Clar}.
In the group theory language the transition takes place between nuclei with SU(5) symmetry and nuclei having  SU(3) symmetry. Recently it has been suggested that the critical point corresponds to a new symmetry called X(5) symmetry \cite{Iache04}.
This shape transition has been also studied within the GCSM formalism in ref.\cite{RaFa}. Here we address the question whether this shape transition is reflected in a specific manner by the behavior of the E0 transition amplitude.
To explore this feature we plotted the predicted as well as the experimental $\rho(E0)$ values as function of $A^{2/3}$, in Fig.6. Note that the experimental results exhibit, indeed,  a maximum for A=154. The theoretical results have been interpolated by a fourth order polynomial which presents an inflexion point for the critical value of A. Thus, we may say that the shape transition is reflected by the fact that an inflexion point shows up in the behavior of the transition amplitude. The bump seen at $^{154}$Gd cannot be obtained in the present model since $k_p$ is depending linearly on the deformation parameter which is varying smoothly with the atomic number. Actually it is hard to say whether the behavior of the energy ratio $E_4^+/E_2^+$ is the most suited  criterion
for deciding whether a phase transition is taking place or not. In the Gd case the {\it critical} nucleus $^{154}$Gd has a static quadrupole deformation of about 0.25 while for  the preceding isotope $\beta_2\approx 0.19$. Therefore the fact that 
$\rho(E0)$ for $^{154}$Gd is larger than the values corresponding to the neighboring isotopes might be determined not by
the transition from a spherical to deformed shapes, which is not the case as we mentioned before, but by another cause. As a matter of fact in Ref.\cite{RaFa}, the analysis of the Hamiltonian structure coefficients as function of the atomic mass suggests that a phase transition is possible to take place in $^{152}$Gd and not $^{154}$Gd. The bump which is seen for $^{154}$Gd could be explained by adding the neutron contribution to the E0 transitions. Indeed the data from the last column of Table II, which correspond to the method c) of fixing $k_p$, indicate a bump for $^{156}$Gd. 

Concerning the model capacity of describing the nuclear shape coexistence, we have derived a compact formula relating the $E0$ transition amplitude with the mixing coefficient of the states describing the nuclear system with different shapes. Indeed, calculating the average value of the static quadrupole moment with the projected state $0^+_{\beta}$, one finds a value which is different from that corresponding to $0^+_g$. Therefore the strength of the $E0$ transition may give information about the structure of the states $0^+_{II}$, namely whether it involves the component $0^+_g$.  
As mentioned before, mixing states corresponding to different shapes may be used to improve the description of the E0 transitions.
In this context we invoke some microscopic calculations for nuclear charge radii and electric quadrupole moments \cite{Pomor} which predict for the even isotopes of $Gd$ oblate equilibrium shapes whose energies are by about 1.5  to 5 MeV larger than the prolate shape energies. The question we would like to address  is {\it how large should be the mixing amplitude of the  states $0^+_g$ and $0^+_{\beta}$, in order that the calculated
transition amplitude $\rho_{I,II}$ be equal to the corresponding experimental data ?} Solving Eq. (\ref{eqforla})
for $^{152,156,158}$Gd, one finds, for $\lambda$, the values: (0.365; 0.03), (0.627;0.002), (0.737;0.016), respectively.
As for $^{154}$Gd, Eq.(4.11) has no real solution. Thus, there is no mixing between 
$0^+_g$ and $0^+_{\beta}$ in the critical nucleus $^{154}$Gd.

Several authors consider $^{74}$Kr as an example of isotope exhibiting shape coexistence. A microscopic calculations using Skyrme (Sk3) and a modified version of Skyrme interaction (SG2), due to Van Giai and Sagawa \cite{Giai}, indicate two equilibrium shapes  with close energies \cite{Sarr}.  Indeed, Sk3 calculation predicts a prolate ground state and an oblate state with an excitation energy of about 0.5 MeV. On the other hand the use of SG2 interaction leads to an oblate ground state and a prolate excited equilibrium minimum with the energy of about 1MeV. Although the prolate deformation is quite large ($\approx 0.389$)  the first excited $2^+$ is relatively high ($= 423.96 keV$). The deformation parameter which provides a good description of the ground band energies is $d=1.9$. This deformation value determines a transition amplitude
equal to 7.78 $e^2fm^2$.  Taking for the experimental transition amplitude the value $8.76 e.fm^2$, as given in Ref.\cite{Chan}, we found for the mixing amplitude the values $\lambda_{1,2} =0.360;0.031$.

\renewcommand{\theequation}{6.\arabic{equation}}
\setcounter{equation}{0}
\section{Conclusion}
\label{sec:level 6}
In the previous Sections we described the results obtained within the GCSM model concerning the charge density as well as the monopole transition amplitude. The main results can be summarized as follows.

The expectation values of the scalar charge density on the unprojected and angular momentum projected states were evaluated 
at different q values.
Angular momentum projection effects are unnoticeable for $J=0$, irrespective of deformation. For larger angular momentum (J=20,
as an example) projection gives sizable effects for small values of deformation. This agrees with  results of microscopic studies by Zaringhalam and Negele \cite{Zari}.

 Concerning the $E0$ transition amplitude, a  quite good  agreement with  data was obtained for $k_p=2.5d$ for most nuclei considered. For this value the unprojected ground state is an eigenstate of the charge density operator in  the low momentum regime. A discrepancy is obtained for $^{168}$Er and $^{174}$Yb,
 where the value $k_p=5d$, which obeys the stability equation for the charge of the system when the transfer momentum is small, seems to be more suitable. Employing for $k_p$ the values  corresponding to a realistic description of both, the M1 and E2 properties \cite{Rad3,Rad4}, one obtains for the monopole transition amplitudes  values which are  larger than the experimental data by a factor 2.6 to 5. 

These large deviations are interpreted as being caused by the different responses of the nuclear quadrupole coordinates to the interaction with the transverse and longitudinal components of an external electromagnetic field, respectively.  The fact that $k_p$ obtained here and in Ref.\cite{Rad3} are different is not an inedit puzzle. We recall that the nuclear deformation obtained by fitting the hydrodynamic moment of inertia is very different from the one obtained by fitting the reduced probability for the transition $0^+\to 2^+$.

We also addressed the questions whether the $E0$ transition amplitude might bring information about shape transitions as well as about shapes coexistence.
Indeed, we notice that for $^{154}$Gd, where the critical point for the spherical to deformed shape transition is met, the transition amplitude $\rho(E0)$ exhibits an inflexion. We suspect that a better method of fixing the parameter $k_p$ would yield for the mentioned function a maximum  value for the critical value $A=154$, as actually happens for the experimental data.

We derived a compact formula relating the $E0$ transition amplitude with the mixing coefficient of the states describing the nuclear system with different shapes. This formula may be used to calculate the mixing amplitude once we know the transition amplitude. Reversely, assuming the state mixing one can calculate the E0 transition amplitude relating the two independent mixed states. Actually we made the option for the second manner of using the mentioned formula and applied it for the isotopes of Gd and $^{74}$Kr. We pointed out that the mixing amplitude could serve as signature for the shape transition in
the Gd isotopes. Indeed for $^{154}$Gd we found out that there is no real solution for the mixing amplitude.

Very recently, a paper addressing similar issues, but within
the IBA approach, showed up \cite{Isaak}. The expression of $\rho(E0)$ obtained therein depends on proton and neutron effective charges and a parameter which is fixed by fitting the peaks of isotope shift. Due to the way of fixing the two parameters the bump seen for $^{154}$Gd is  reproduced by the calculation of Ref.\cite{Isaak}.
It is worth mentioning that, by contrast, we don't use effective charges and we do not have a parameter which is fixed by fitting the peaks in the isotope shift, which is a phenomenon closely related to the E0 properties. Despite these differences, both descriptions provide results for the $E0$ transitions in $Gd$ isotopes, which agree with the corresponding experimental data. The agreements  obtained in the two descriptions are of similar quality. Exception is for  the bump which shows up at $^{154}$Gd. Indeed, our calculations predict an inflexion point rather than a bump. For comparison the two sets of results, ours and those of Ref.\cite{Isaak} for
$Gd$ isotopes  were represented on the same graph, in Fig. 6.

Finally, one could assert that the GCSM model provides results for the diagonal matrix elements of the charge density
 operator,  which are consistent with the microscopic studies \cite{Coop,Zari,Anto}. Also, the model is able to realistically describe the $E0$ transitions for the eleven nuclei considered in the present paper. Comments are made upon the signatures of nuclear shape phase transition as well of shape coexistence which might be found in the monopole charge transition.

{\bf Acknowledgments.} A.A.R. wants to thank UCM-GRUPO SANTANDER for financial support of his visit at Complutense University of Madrid where part of this work has been performed.
Two of us (E.M.G. and P. S.) acknowledge MEC (Spain) for financial support (FIS2005-00640).
This work was also supported  by the Romanian Ministry for Education and Research under the contract PNII, No. ID-33/2007.


\begin{references}
\bibitem{Coop} T.Cooper {\it et al.}, Phys. Rev. {\bf C 13} (1976) 1083.
\bibitem{Zari}A. Zaringhalam and J. W. Negele, Nucl. Phys., {\bf A288} (1977) 417.
\bibitem{Vill} F. Villars, in {\it Many-Body Description of Nuclear Structure and Reactions} (C. Bloch, ed.), Academic Press, New York, 1966.
\bibitem{Elvi} E. Moya de Guerra, Phys. Rep. {\bf 138}, No. 6 (1986) 293.
\bibitem{Sarr} P. Sarriguren, E. Moya de Guerra, A. Escuderos and A. C. Carrizo, Nucl. Phys. {\bf A 635} (1998) 55.
\bibitem{Les} S. R. Lesher {\it et al.,} Phys. Rev. {\bf C 66} (2002) 051305 (R)
\bibitem{Nic} N. Lo Iudice, A. V. Sushkov and N. Yu. Shirikova, Phys. Rev. {\bf C 70} (2004) 064316.
\bibitem{Iudi} N. Lo Iudice and F. Palumbo, Phys. Rev. Lett. {\bf 41},(1978) 1532; G. De Francheschi, F. Palumbo and N. Lo Iudice, Phys. Rev. {\bf C29} (1984) 1496.
\bibitem{Rad1}A. A. Raduta, V. Ceausescu, A. Gheorghe and R. Dreizler, Phys. Lett. {\bf B 121} (1981) 1; Nucl. Phys. {\bf A 381}
(1982) 253.
\bibitem{Rad2}A. A. Raduta, A. Faessler and V. Ceausescu, Phys. Rev. {\bf C 36} (1987) 439.
\bibitem{Rad3}A. A. Raduta, I. I. Ursu and D. S. Delion, Nucl. Phys. {\bf A 475} (1987) 439.
\bibitem{Rad4}A. A. Raduta and D. S. Delion, Nucl. Phys. {\bf A 491} (1989) 24.
\bibitem{Iud}N. Lo Iudice, A. A. Raduta and D. S. Delion, Phys. Rev. {\bf C50} (1994) 127
\bibitem{Rad5}A. A. Raduta, A. Gheorghe and M. Badea, Z.  Phys. {\bf A283} (1977) 79
\bibitem{Rad6}A. A. Raduta and C. Sabac, Ann Phys. (NY) {\bf 148} (1983) 1.
\bibitem{Rad7}A.A.Raduta, Coherent State Model for several collective interacting bands, in Recent Developments in Nuclear Physics,1 (2004):1-70, ISBN:81-7895-124-X.
\bibitem{Bohr}A.Bohr and B. Mottelson, Nuclear Structure, Volume I: Single particle Motion, World Scientific,1998, p.383.
\bibitem{Lang} J. Lange, Krishna Kumar and J. H. Hamilton, Rev. Mod. Phys. {\bf 54} (1982) 119.
\bibitem{Wood} J. L. Wood, E. F. Zganjar, C. De Coster and K. Heyde, Nucl. Phys. {\bf A651} (1999) 323.
\bibitem{Gar} P. E. Garrett, J. Phys. G: Nucl. Part. Phys. {\bf 27} (2001) R1-R22.
\bibitem{Cacsi} Z. Gacsi {\it et al.} Phys. Rev. {\bf 64} (2001) 047303.
\bibitem{Kibe} T. Kibedi and R. H. Spear, Atomic Data and Nuclear Data Tables {\bf 89} (2005) 77.
\bibitem{Wimm} K.Wimmer {\it et al.}arXiv:0802.2514v1[nucl-ex] 18 Feb 2008.
\bibitem{Anto} A. N. Antonov, {\it et al}., Phys. Rev. {\bf C 72} (2005) 044307.
\bibitem{Clar} R. M. Clark {\it et al}., Phys. Rev. {\bf C 68} (2003) 037301.
\bibitem{Iache04} F. Iachello, Phys. Rev. Lett. {\bf 87} (2001) 052502.
\bibitem{RaFa} A. A. Raduta and Amand Faessler,J. Phys. G: Nucl. Part. Phys. {\bf 31} (2005) 873.
\bibitem{Pomor} Bozena Nerlo-Pomorska and Beata Mach, Atomic Data and Nuclear Data Tables {\bf 60}, (1995) 287.
\bibitem{Giai} N. Van Giai and H. Sagawa, Phys. Lett. {\bf B 106} (1981) 379.
\bibitem{Chan} C. Chandler {\it et al.} Phys. Rev. {\bf C56} (1997) R2924.
\bibitem{Isaak} S. Zergine, P. Van Isacker, A. Bouldjedri, S. Heinze, Phys. Rev. Lett. {\bf 101}, (2008) 022502.
\end{references}
\end{document}